\documentclass{article}
\usepackage{graphicx}

\begin{document}

\title{\textbf{ Local Fields without
Restrictions on the Spectrum of 4-Momentum Operator
and Relativistic Lindblad Equation}}
\author{M.A. Kurkov \footnote{
High Energy and Elementary Particles Physics Department,
The Faculty of Physics, Saint-Petersburg State University,
e-mail: katok86@mail.ru}
\and V.A. Franke\footnote{
High Energy and Elementary Particles Physics Department,
The Faculty of Physics, Saint-Petersburg State University,
e-mail: franke@pobox.spbu.ru}}

\date{}
\maketitle

\begin{abstract}

Quantum theory of Lorentz invariant  local scalar fields
without restrictions on 4-momentum
spectrum is considered. The mass spectrum
may be both discrete and continues and the square of mass
as well as the energy may be positive or negative.
Such fields can exist as part of a hidden matter
in the Universe if they interact
with ordinary fields very weakly.
Generalization of Kallen-Lehmann representation
for propagators of these fields is found. The considered
generalized fields may violate $CPT$- invariance. Restrictions
on mass-spectrum of $CPT$-violating fields are found.
Local fields that annihilate vacuum state and violate $CPT$-
invariance are constructed in this scope.
Correct local relativistic generalization of Lindblad equation
for density matrix is written for such fields.
This generalization is particulary needed to describe the evolution
of quantum system and measurement process in a unique way.
Difficulties arising when the field annihilating the vacuum interacts
with ordinary fields are discussed.

\end{abstract}
\textbf{keywords:} tachyons, $CPT$-violation, collapse of the state vector, \\
Lindblad equation,
renormalizability

\section*{1.~~Introduction.}

It is known that there exists a lot of hidden mass in the Universe,
and its interaction with usual matter is very weak.
Such weakness of interaction allows us
to suppose without contradiction with experiment,
that hidden mass contains fields, which do not satisfy usual restrictions imposed on
the spectrum of 4-momentum. In this work we consider
the properties of scalar nonhermitian fields of such type and discuss
applications of these fields. Although the mentioned
fields are of special interest in cosmology,
we consider as the first step the case of flat space-time.
With the exception of the requirements for the spectrum of 4 momentum, which we refuse,
all other postulates of Quantum Field Theory and in particular
the property of local commutativity are kept in the following consideration.
We consider scalar local fields both with discrete and
continues mass spectrum. These fields may have positive or negative  square of mass
and positive or negative energy in the case of positive square of mass.
Let us remark, that some time ago a theory of tachyonic fields (i.e.
fields with negative square of mass) was proposed based on the assumption
that only particles with positive energy can be created [1].
Such tachyonic scalar fields fulfill anticommutation relations.
No field, annihilating the vacuum exists in this model. On the
contrary we suppose that states with arbitrary sign of energy
can be created, assuming for consistency with experiment
that the interaction with conventional matter is extremely weak.
This permits us to use commutation relations between scalar fields only.
In such a way we get the local field annihilating the vacuum and apply
it in subsequent consideration.

In the second section the generalization of Kallen-Lehmann representation
for the propagator of nonhermitian scalar
field is deduced under assumptions described above.
It is clarified that CPT- violation in local Lorentz invariant theory
may take place, if we abandon ordinary
requirements for the spectrum of 4-momentum.
At the cost of such violation it is possible to construct nonzero
local field annihilating the vacuum state.
We shall use \emph{\textbf{free}} field of such type in the following consideration
for particular applications.

In the third chapter it is shown, how generalized free field
with arbitrary spectrum of 4-momentum can be constructed,
if its propagator coincides with the one,
given by general expression found in section 2.

In the forth section the local relativistic generalization of Lindblad equation
for density matrix is introduced as an example of usage of the
fields mentioned above.
In  nonrelativistic  Quantum Mechanics such equation allows, in particular,
to describe in a unique way the evolution of a quantum system over time
as well as the measurement process.
Generalization of quantum theory by the passage from $Schr\ddot{o}dinger$
equation to Lindblad equation is especially advisable in cosmology
to describe the early Universe. In this case it is meaninglessly to speak
about any external devices making measurements over the Universe.
That's why one needs an equation which provides spontaneous transformation
of superposition of macroscopically different states of the Universe
into one of them.
In order to apply Lindblad equation for this purpose,
one has to write its relativistic
generalization  for flat space-time, and then pass
to Riemann-space of gravitational theory.
Here we consider only the first part of this problem, and assume the space-time
to be flat.
Trying to construct a local
relativistic generalization of Lindblad equation,
using only  fields with conventional spectrum of 4-momentum operator,
one meets irresistible ultraviolet
(u.v. below) divergencies [2].
However, in section 4 it is shown that such generalization exists
for the free field annihilating the vacuum.
It should be noted, that the authors of the work [3] with an eye to
describe the evolution of nonrelativistic quantum system and measurement
process in a unique way,
used stochastic $Schr\ddot{o}dinger$ equation, which
corresponds to  specified Lindblad equation.
These authors emphasize, that the
usage of corresponding Lindblad equation does not mean yet the
passage to realistic description which provides
all macroscopic quantities with definite values.
It is connected with the fact, that
the density matrix $\rho$ can be presented as a sum of
projectors onto pure states in different ways and the Lindblad equation
does not give an instruction which of them should be chosen.
Nevertheless,
if the density matrix $\rho(t)$ satisfies the Lindblad equation
and can be presented as a sum of projectors
onto macroscopically definite states
at every moment of time $t$ at least in one way, then
apparently one can construct
stochastic $Schr\ddot{o}dinger$ equation which generates this Lindblad equation.
So we describe the Lindblad equation only.

In the fifth section we consider u.v. divergencies arising
when one introduces the interaction between
the field annihilating the vacuum and ordinary fields.
In the investigated examples these divergencies have unusual nature and
can not be removed by renormalization.
In particular, it inhibits to use
the considered local Lindblad equation for description of
collapse of the state vector of usual fields.

\section*{2.~~Kallen-Lehmann representation at
arbitrary spectrum of 4-momentum operator.}
For the sake of simplicity we shall
consider the case of nonhermitian scalar field $\varphi(x)$
$(\varphi^{\dag}(x) \neq \varphi(x))$ only.
It is more interesting than a hermitian one,
that could be easily described in a similar way. We assume that the space-time
is flat with the metric
$g_{\mu\nu}=diag(+1,-1,-1,-1)$ and the theory is Lorentz and
translation invariant.  In the whole section 2 the
Heisenberg representation is used, and the symbol $|0\rangle$
means physical (Heisenberg) vacuum state,
normalized by the condition $\langle0|0\rangle=1$.
We assume that the $|0\rangle$ is Lorentz-
and translation invariant. The field $\varphi(x)$ is required to
satisfy the locality conditions:
$$\left[\varphi(x),\varphi(x')\right]=0~~~at~~~(x-x')^2<0, \eqno{(2.1a)}$$
$$\left[\varphi(x),\varphi^{\dag}(x')\right]=0~~~at~~~(x-x')^2<0, \eqno{(2.1b)}$$
where $(x)^2 = g_{\mu\nu}x^\mu x^\nu$; $\mu,\nu,...=0,1,2,3.$
For simplicity the theory is considered to be invariant
under the global phase transformations
$$\varphi\rightarrow e^{i\alpha}\varphi, \eqno(2.2)$$
where $\alpha$ is an arbitrary real constant. The vacuum
$|0\rangle$ is assumed to be invariant under (2.2). Therefore
$$\langle 0 |\varphi(x)\varphi(x')|0\rangle = 0. \eqno{(2.3)}$$
Due to (2.3) the casual and the retarded Green functions of  $\varphi(x)$ and
$\varphi(x')$ are also equal to zero. The following
Wightman functions and casual Green function are nonzero :
$$u(x)\equiv \langle 0 |\varphi(x)\varphi^{\dag}(0)|0\rangle, ~~~~~
w(x)\equiv \langle 0 |\varphi^{\dag}(x)\varphi(0)|0\rangle, \eqno(2.4)$$
$$G(x)\equiv \langle 0 |T\left\{\varphi(x)
\varphi^{\dag}(0)\right\}|0\rangle, \eqno(2.5)$$
where
$T\left\{\varphi(x)\varphi^{\dag}(y)\right\}
=\theta(x^0-y^0)\varphi(x)\varphi^{\dag}(y)
+\theta(y^0-x^0)\varphi^{\dag}(x)\varphi(y)$ and $\theta(a)=1$ at $a>0,
\theta(a)=0$ at $a<0$. Because of translation invariance
$\langle 0 |\varphi(x)\varphi^{\dag}(y)|0\rangle = u(x-y)$,
and likewise for other two-point functions.
The problem under consideration is to find general representation for $G(x)$
without any limitations for the spectrum of 4-momentum operator , i.e. to
generalize Kallen-Lehmann representation.

For every function $\xi(x)$ we introduce the Fourier transform $\widetilde{\xi}(k)$,
assuming that
$$\widetilde{\xi}(k)=\int d^4x e^{ikx}\xi(x),
~~~~~ \xi(x)=\frac{1}{(2\pi)^4}\int d^4k e^{-ikx}\widetilde{\xi}(k),\eqno{(2.6)}$$
where $kx\equiv k^0x^0-k^i x^i \equiv k^0 x^0 - \vec k\vec x, ~i=1,2,3$.
Because of Lorentz invariance, the
Fourier-transforms $\widetilde{u}(k)$,$~\widetilde{w}(k)$
of the quantities $u(x)$,  $~w(x)$, defined by the formulas (2.4), can depend only on
$k^2$ when $k^2<0$, and on $k^2$ as well as
on the sign of $k^0$ when $k^2 \geq 0$, so
\footnote{Further
in those places where it does not cause misunderstandings we shall sometimes
write $\alpha(k)$ and $\beta(k)$ instead of  $\alpha(k^2,\theta(k^2)sgn(k_0))$ and
$\beta(k^2,\theta(k^2)sgn(k_0))$ for short.}
$$\widetilde{u}(k)=\alpha(k^2,\theta(k^2)sgn(k_0)),
~~~~~\widetilde{w}(k)=\beta(k^2,\theta(k^2)sgn(k_0)).\eqno(2.7)$$
Now we introduce  complete set of eigenvectors $|k\rangle$ of 4-momentum
operator $P^\mu$:
$$P^\mu|k\rangle=k^\mu|k\rangle. \eqno(2.8)$$
The states $|k\rangle$ are normalized in such a manner,
that the unity operator $I$ has the following form
$I=\int d^4k|k\rangle\langle k|$. Due to translation invariance
$$\varphi(x)=e^{iPx}\varphi(0)e^{-iPx},~~\varphi^{\dag}(x)
=e^{iPx}\varphi^{\dag}(0)e^{-iPx}. \eqno(2.9)$$
Based on the formulas (2.4), (2.6), (2.7), (2.9) and the relations
$$P^\mu|0\rangle=0,~~\langle0|P^\mu=0 \eqno(2.10)$$
one finds, that
$$\widetilde{u}(k)\equiv\alpha(k^2,\theta(k^2)sgn(k^0))
=(2\pi)^4\langle 0|\varphi(0)|k\rangle\langle k
|\varphi^{\dag}(0)|0\rangle, \eqno(2.11)$$
$$\widetilde{w}(k)\equiv\beta(k^2,\theta(k^2)sgn(k^0))
=(2\pi)^4\langle 0|\varphi^{\dag}(0)|
k\rangle\langle k |\varphi(0)|0\rangle. \eqno(2.12)$$
This implies, that
$$\alpha(k^2,\theta(k^2)sgn(k_0))\geq 0, ~~~~~
\beta(k^2,\theta(k^2)sgn(k_0))\geq 0.\eqno(2.13)$$
One sees, that nonnegative functions $\alpha$ and $\beta$ are defined by
4-momentum spectrum of physical states of the theory. Further we show, that casual
Green function $G(x)$ can be expressed
through $\alpha$ and $\beta$, and it turns out,
that this quantities are subordinated to some conditions due to (2.1b).

Let us define the function
$$F(x)=i\langle0|[\varphi(x),\varphi^{\dag}(0)]|0\rangle, \eqno(2.14)$$
and also retarded and advanced Green functions:
$$F_r(x)=\theta(x^0)F(x),~~~~~ F_a(x)=-\theta(-x^0)F(x),\eqno(2.15)$$
so that
$$F(x)=F_r(x)-F_a(x). \eqno(2.16)$$
Owning to locality condition (1) the functions
$F_r(x),~F_a(x)$ are  Lorentz invariant, and
$$F(x)=0 \quad\mbox{at}\quad (x)^2<0, \eqno(2.17)$$
$$F_r(x)\qquad\mbox{is nonzero  only if}\qquad(x)^2\geq 0,~ x^0\geq 0, \eqno(2.18)$$
$$F_a(x)\qquad\mbox{is nonzero  only if}\qquad(x)^2\geq 0,~ x^0\leq 0. \eqno(2.19)$$
Then we form Fourier transforms
$\widetilde{F}(k),~\widetilde{F_r}(k),~\widetilde{F_a}(k)$
of functions $F(x)$, $F_r(x)$, $F_a(x)$ according to (2.6).
Because of (2.4), (2.7), (2.14)
$$\widetilde{F}(k)=i(\widetilde{u}(k)-\widetilde{w}(-k))=
i(\alpha(k^2,\theta(k^2)sgn(k^0))-\beta(k^2,-\theta(k^2)sgn(k^0)).\eqno(2.20)$$

So far, it was assumed that $k^\mu$ is a real vector.
Now we shall investigate the analytical continuation
into the region of complex $k^\mu$, i.e. we
allow the vector $k^\mu$ in the formulas
$$\widetilde{F_r}(k)=\int d^4x e^{ikx}F_r(x), ~~~~~
\widetilde{F_a}(k)=\int d^4x e^{ikx}F_a(x) \eqno(2.21)$$
to take complex values. We shall assume, that
$F(x)$, $~F_r(x),$ $~F_a(x)$ are distributions of slow growth (from $S'$ ). Then
$\widetilde{F_r}(k),~\widetilde{F_a}(k)$ are analytical functions of the
argument $k^\mu$ at those its values for
which the first integral (2.21), or correspondingly
the second, exists.\footnote{Here are applicable all the considerations
described in the book [4]
 with changing $x_{\mu}$ to $k_{\mu}$ and vice versa.}

According to  (2.18), (2.19)
$$\widetilde{F_r}(k)\qquad\mbox{is analytical at}
\qquad (Im~k)^2>0,~Im~k^0>0, \eqno(2.22)$$
$$\widetilde{F_a}(k) \qquad\mbox{is analytical at}
\qquad (Im~k)^2>0,~Im~k^0<0. \eqno(2.23)$$
But due to Lorentz invariance, the functions $\widetilde{F_r}$ and $\widetilde{F_a}$
can depend only on $k^2$ in those region, where they are analytical.  The region
of analyticity contains those $k^2$, which can be expressed
through $k^\mu$ under the condition (2.22) (correspondingly (2.23))at least in a one way.
It is easy to check, that conditions (2.22), (2.23) can be satisfied
at every $k^2$, expect those ones, for which
$$Im(k^2)=0,~~~~~Re(k^2)\geq 0. \eqno(2.24)$$
So, in the region of analyticity we have
$$\widetilde{F_r}(k)=f_r(k^2),~~~~~\widetilde{F_a}(k)=f_a(k^2),  \eqno(2.25)$$
and the functions $f_r(k^2)$,$~f_a(k^2)$
are analytical in a whole complex $k^2$-plane,
probably with the exception of the positive real axis.
The functions $\widetilde{F_r}(k)$, $~\widetilde{F_a}(k)$
at $Im~k^\mu=0,~~Re(k^2)\geq 0$
can be expressed through $f_r(k^2)$, $~f_a(k^2)$, going to the limit
from those $k^\mu$, for which $\widetilde{F_r}(k),$
(correspondingly $\widetilde{F_a}(k)$)
is analytical. Taking into account (2.22),(2.23),
it is easy to conclude,
that at real $k^\mu$
$$\widetilde{F_r}(k)=\lim_{\epsilon\rightarrow +0}
f_r(k^2+i\epsilon~sgn(k_0)),~~~~~
\widetilde{F_a}(k)=\lim_{\epsilon\rightarrow +0}
f_a(k^2-i\epsilon~sgn(k_0)). \eqno(2.26)$$
In the following we consider $k^\mu$
to be real and write the formulas (2.26) without
the sign of limit.
Due to the definition (2.14)
$$F(-x)=-F^*(x)\eqno(2.27)$$
(*-is a sign of complex conjugation).
Thus,  using (2.15), we find, that
$$\widetilde{F_a}(k)=\widetilde{F_r}^*(k),~~~~~
\widetilde{F}(k)=-\widetilde{F}^*(k),\eqno(2.28)$$
and because of (2.26)
$$f_a(k^2-i\epsilon~sgn~k_0)=f_r^*(k^2+i\epsilon~sgn~k_0)
=f_r^*((k^2-i\epsilon~sgn~k_0)^*).\eqno(2.29)$$
This means, that analytical functions $f_a(s)$, $f_r(s)$  satisfy the condition
$$f_a(s)=f_r^*(s^*)\eqno(2.30)$$
at every complex $s$.
In the following we shall write $f(s)$ instead of $f_r(s)$.
So according to (2.26), (2.29)
$$\widetilde{F_r}(k)=f(k^2+i\epsilon~sgn(k_0)),~~~~~
\widetilde{F_a}(k)=f^*(k^2+i\epsilon~sgn~(k_0)),  \eqno(2.31)$$
$$\widetilde{F}(k)=f(k^2+i\epsilon~sgn(k_0))-f^*(k^2+i\epsilon~sgn(k_0))
=2Im~f(k^2+i\epsilon~sgn(k_0)). \eqno(2.32)$$
Thus due to (2.20)
$$Im~f(k^2+i\epsilon~sgn(k_0))=\frac{1}{2}\Bigl(\alpha(k^2,\theta(k^2)sgn(k^0))
-\beta(k^2,\theta(k^2)sgn(k^0))\Bigr).\eqno(2.33)$$

The analytical function $f(s)$ has a cut along the positive real axis,
where the relation (2.33) is true. So $f(s)$ can be reconstructed
\footnote{Long time ago Yuriy Petrovich Scherbin, who is no more with us,
taught one of the authors (V. A. Franke)
the following mathematical procedure. }
using the functions $\alpha$ and $\beta$, if they are defined at $k^2\geq 0$.
We shall do it at first imposing restrictions on the function $f(s)$ strong enough,
to provide its uniqueness, and later we shall discuss the arbitrariness,
caused by weakening these restrictions.
Let us introduce into consideration the function $\sqrt{-s}$
of complex variable $s$ and
let us define the branch of the square root as follows:
if $ s=  |s|e^{i\mu},~ - \pi< \mu < \pi $, then
$$\sqrt{-s}=-i\sqrt{|s|}e^{\frac{i\mu}{2}}
\quad\mbox{at}\quad \mu\geq 0,\qquad\sqrt{-s}= i\sqrt{|s|}
e^{\frac{i\mu}{2}}\quad\mbox{at}\quad\mu\leq 0.\eqno(2.34)$$
Thus at real $k^2$ in the limit $\epsilon\rightarrow 0$ with $\epsilon>0$
$$\sqrt{-(k^2\pm i\epsilon)} = \mp i\theta(k^2)\sqrt{k^2}
+ \theta(-k^2)\sqrt{|k^2|}.\eqno(2.35)$$
Let us define the functions
$$\xi_+(s)\equiv\frac{1}{2}\left(f(s)+f^*(s^*)\right),~~~~~
\xi_-(s)\equiv -\frac{i}{2\sqrt{-s}}\left(f(s)-f^*(s^*)\right),\eqno(2.36).$$
Then because of (2.34)
$$\xi^*_{\pm}(s^*)=\xi_{\pm}(s).\eqno(2.37)$$
Furthermore the functions $\xi_{\pm}(s)$ have the same analytical properties,
as the function $f(s)$. Let us assume,
that, when $|s| \rightarrow \infty $, the functions $\xi_{\pm}(s)$ decrease
not slower than $|s|^{-\alpha}$, but when $s \rightarrow 0$
they increase not faster than $|s|^{-1+\alpha}$, where $\alpha > 0$.
Then due to (37) the following dispersion relations are true:
$$\xi_{\pm}(s)=\frac{1}{\pi}\int\limits_{0}^{\infty}ds'\frac{1}{s'-s}
Im~\xi_{\pm}(s'+i\epsilon),\eqno(2.38)$$
so according  to (2.35), (2.36) at $s\geq0$
$$Im~\xi_+(s+i\epsilon)=\frac{1}{2}(Im
~f(s+i\epsilon)-Im~f(s-i\epsilon)),\eqno(2.39)$$
$$Im~\xi_-(s+i\epsilon)=\frac{1}{2\sqrt{s}}
(Im~f(s+i\epsilon)+Im~f(s-i\epsilon)).\eqno(2.40)$$
Taking into account the relations (2.33),(2.38),(2.39),(2.40), we conclude that
$$\xi_{+}(s)=\frac{1}{4\pi}\int\limits_{0}^{\infty}ds'\frac{1}{s'-s}
\Bigl(\alpha(s',1)+\beta(s',1)-\alpha(s',-1)-\beta(s',-1)\Bigr),\eqno(2.41)$$
$$\xi_{-}(s)=\frac{1}{4\pi}\int\limits_{0}^{\infty}ds'\frac{1}{(s'-s)\sqrt{s'}}
\Bigl(\alpha(s',1)-\beta(s',1)+\alpha(s',-1)-\beta(s',-1)\Bigr).\eqno(2.42)$$
Finally, because of (2.36)
$$f(s)=\frac{1}{4\pi}\int\limits_{0}^{\infty}ds'\frac{1}{s'-s}
\biggl(\alpha(s',1)+\beta(s',1)-\alpha(s',-1)-\beta(s',-1)\biggr)+\eqno(2.43)$$
$$+\frac{i\sqrt{-s}}{4\pi}\int\limits_{0}^{\infty}ds'\frac{1}{(s'-s)\sqrt{s'}}
\biggl(\Bigl(\alpha(s',1)-\beta(s',1)\Bigr)+
\Bigl(\alpha(s',-1)-\beta(s',-1)\Bigr)\biggr).$$
Here $\alpha(s',1)=\alpha(k^2,\theta(k^2)sgn(k^0))~$at
$ ~k^2=s'\geq 0$, $\theta(k^2)sgn(k^0)=1$
and similarly for $\beta(s',1)$,$~\alpha(s',-1)$,$~\beta(s',-1).$

Let us consider the equality (2.43) from the point of view of $CPT$-transformation.
At such transformation the matrix element
 $\langle 0|\varphi(x)\varphi^{\dag}(0)|0\rangle$  goes into
 $\langle 0|\varphi^{\dag}(0)\varphi(-x)|0\rangle~$
 $=\langle 0|\varphi^{\dag}(x)\varphi(0)|0\rangle$ and vice versa,
or according to (2.4), (2.6),(2.7)
$$\alpha(k^2,\theta(k^2)sgn(k^0))\quad\mbox{and}\quad\beta(k^2,\theta(k^2)sgn(k^0))
\quad\mbox{pass into each other}~.\eqno(2.44)$$
So, the first term in the
right hand side of the equality (2.43) is $CPT$- invariant,
while the second one is not $CPT$- invariant.
Assuming that $s=k^2<0$ in the formula (2.43),  taking the
imaginary part of the function $f(k^2)$ and using (2.33)
at $k^2<0$, we obtain the relation
$$\theta(-k^2)\Bigl(\alpha(k^2,0)-\beta(k^2,0)\Bigr)=
\theta(-k^2)\frac{\sqrt{-k^2}}{2\pi}\times$$
$$\times\int\limits_{0}^{\infty}ds'\frac{1}{(s'-k^2)\sqrt{s'}}
\biggl(\Bigl(\alpha(s',1)-\beta(s',1)\Bigr)+
\Bigl(\alpha(s',-1)-\beta(s',-1)\Bigr)\biggr).\eqno(2.45)$$
One sees, that $CPT$- noninvariant
parts of the spectrum are not arbitrary,
but they are coupled by the relation (2.45).
This is a result of the locality of the theory.
In the equality (2.45) the expression $(\alpha(s',1)-\beta(s',1))$ describes the
$CPT$- violation in the spectrum of states with $s'=k^2\geq 0$ and positive energy,
the difference $(\alpha(s',-1)-\beta(s',-1))$
- in the spectrum  of states with nonnegative $s'=k^2$, but with
negative energy and, finally,
$\theta(-k^2)(\alpha(k^2,0)-\beta(k^2,0))$- in the spectrum of tachyonic
states ($k^2<0$). Let us notice, that $CPT$-invariant parts of the spectrum
$~\theta(+k^2)(\alpha(k^2,1)+\beta(k^2,1))$,
$~\theta(+k^2)(\alpha(k^2,-1)+\beta(k^2,-1))$,
$~\theta(-k^2)(\alpha(k^2,0)+\beta(k^2,0))$
can be defined arbitrary.

Now let us consider the propagator $G(x)$, defined by the equality (2.5).
Using the relations (2.4), (2.5), (2.14), (2.15)
one sees, that $G(x)$ might be written,in particular,
 in two ways:
$$G(x)=F_r(x)+iw(-x)=F_a(x)+iu(x).\eqno(2.46)$$
Using the equalities (2.31), (2.7) we obtain for the Fourier transform
$$\widetilde{G}(k)=f(k^2+i\epsilon~sgn(k^0))+i\beta(k^2,-\theta(k^2)sgn(k^0)) =$$
$$= f^*(k^2+i\epsilon~sgn(k^0))+i\alpha(k^2,\theta(k^2)sgn(k^0)),\eqno(2.47)$$
i.e. in accordance with (2.43)
$$\widetilde{G}(k)=\frac{1}{4\pi}\int\limits_{0}^{\infty}ds'
\frac{1}{s'-k^2-i\epsilon~sgn(k^0)}
\biggl(\Bigl(\alpha(s',1)+\beta(s',1)\Bigr)-
\Bigl(\alpha(s',-1)+\beta(s',-1)\Bigr)\biggr)+$$
$$+\frac{i\sqrt{-k^2-i\epsilon ~ sgn(k^0)}}{4\pi}\int\limits_{0}^{\infty}ds'
\frac{1}{(s'-k^2-i\epsilon~sgn(k^0))\sqrt{s'}}\times \eqno(2.48a)$$
$$\times\biggl(\Bigl(\alpha(s',1)-\beta(s',1)\Bigr)+
\Bigl(\alpha(s',-1)-\beta(s',-1)\Bigr)\biggr)
+i\beta(k^2,-\theta(k^2)sgn(k^0)),$$
or
$$\widetilde{G}(k)=\frac{1}{4\pi}\int\limits_{0}^{\infty}ds'
\frac{1}{s'-k^2+i\epsilon ~sgn(k^0)}
\biggl(\Bigl(\alpha(s',1)+\beta(s',1)\Bigr)-
\Bigl(\alpha(s',-1)+\beta(s',-1)\Bigr)\biggr)+$$
$$-\frac{i\sqrt{-k^2+i\epsilon ~ sgn(k^0)}}{4\pi}\int\limits_{0}^{\infty}ds'
\frac{1}{(s'-k^2+i\epsilon ~sgn(k^0))\sqrt{s'}}\times \eqno(2.48b)$$
$$\times\biggl(\Bigl(\alpha(s',1)-\beta(s',1)\Bigr)+
\Bigl(\alpha(s',-1)-\beta(s',-1)\Bigr)\biggr)
+i\alpha(k^2,\theta(k^2)sgn(k^0)).$$
This is the required generalization of Kallen-Lehmann representation.
Let us rewrite it in another form, which allows us to understand more clearly
the meaning of each term of the sum. Using well known formulas like
$$\frac{1}{x+ i\epsilon}+2\pi i\delta(x)=\frac{1}{x-i\epsilon},\eqno(2.49)$$
one can show, that (we omit simple, but lengthy calculations)
$$\widetilde{G}(k)=\frac{1}{4\pi}\int\limits_{0}^{\infty}ds'
\left(\frac{\alpha(s',1)+\beta(s',1)}{s'-k^2-i\epsilon }
-\frac{\alpha(s',-1)+\beta(s',-1)}{s'-k^2+i\epsilon }\right)+$$
$$+\theta(-k^2)\frac{i}{2}\Bigl(\alpha(k^2,0)+\beta(k^2,0)\Bigr)+\eqno(2.50)$$
$$+\theta(k^2)\frac{\sqrt{k^2}~sgn(k^0)}{4\pi}\int
\limits_{0}^{\infty}\frac{ds'}{\sqrt{s'}}
\left(\frac{\alpha(s',1)-\beta(s',1)}{s'-k^2-i\epsilon }
+\frac{\alpha(s',-1)-\beta(s',-1)}{s'-k^2+i\epsilon }\right).$$
Ordinary Kallen-Lehmann representation [5][6] could be obtained from (2.50), by putting
$\alpha(k^2,-1)=\beta(k^2,-1)=\alpha(k^2,0)
=\beta(k^2,0)=0,~\alpha(k^2,1)=\beta(k^2,1)$.
So only the first term under the sign of the first integral remains.
In the general case
there is the similar term  in the first integral,
which contains $\alpha(s',-1)+\beta(s',-1)$.
It describes the contribution of the states with
$k^2\geq 0,~k^0<0.$ Both of this terms are $CPT$- invariant.
Further, outside the integral  there is a
$CPT$-invariant tachyonic contribution, which contains
$\alpha(k^2,0)+\beta(k^2,0)$. Finally there is $CPT$- noninvariant contribution, given
by the last integral in the formula (2.50).
This contribution is nonzero only if $k^2\geq 0$.
Values of function $\widetilde{G}(k)$ at $k^2<0$ are $CPT$- invariant.
We emphasize that the quantities $\alpha(k^2,1),~\beta(k^2,1),$ $~\alpha(k^2,-1),$
$~\beta(k^2,-1),~\alpha(k^2,0),~\beta(k^2,0)$
are coupled with each other by the relation (2.45), although the difference
$\Bigl(\alpha(k^2,0)-\beta(k^2,0)\Bigr)$ is absent at $k^2<0$ in the formula (2.50).

If there are no restrictions on the spectrum of 4-momentum operator and
the $CPT$- invariance is violated, then
it is possible to introduce nonzero local field
$\varphi(x)$, annihilating the vacuum state i.e. fulfilling the condition
$$\varphi(x)|0\rangle = 0\qquad\mbox{at arbitrary }\qquad x .\eqno(2.51)$$
In this case due to (2.4) and (2.7)
$$\beta(k^2,\theta(k^2)sgn(k^0))=0,\eqno(2.52)$$
and according to the expression (2.48a)
$$\widetilde{G}(k)=\frac{1}{4\pi}\int\limits_{0}^{\infty}ds'
\frac{1}{s'-k^2-i\epsilon~sgn(k^0)}
\Bigl(\alpha(s',1)-\alpha(s',-1)\Bigr)+$$
$$+\frac{i\sqrt{-k^2-i\epsilon ~sgn(k^0)}}{4\pi}\int\limits_{0}^{\infty}ds'
\frac{1}{(s'-k^2-i\epsilon~sgn(k^0))\sqrt{s'}}\times\eqno(2.53)$$
$$\times\Bigl(\alpha(s',1)+\alpha(s',-1)\Bigr). $$
Furthermore because of (2.45)
$$\theta(-k^2)\alpha(k^2,0)=\theta(-k^2)\frac{\sqrt{-k^2}}{2\pi}\times$$
$$\times\int\limits_{0}^{\infty}ds'\frac{1}{(s'-k^2-i\epsilon~sgn(k^0))\sqrt{s'}}
\Bigl(\alpha(s',1)+\alpha(s',-1)\Bigr).\eqno(2.54)$$

It is easy to see from (2.54), that in the case (2.51)
the spectrum of tachyons is continuous
for every $\alpha(s',1)$ and $\alpha(s',-1)$ and covers all
negative real axis $-\infty < k^2< 0$.

Let us do several remarks. Writing the formula (2.38), we assumed, that
the function $\xi_-(s)$ increases at
$s\rightarrow 0$ not faster, then $|s|^{-1+\alpha}$
where $\alpha>0$. Due to the equality (2.36)
it means, that the difference $f(s)-f^*(s^*)$
increases in the limit $s\rightarrow 0$ not
faster then $|s|^{-\frac{1}{2}+\alpha}$.
One could weaken this condition and consider the function
$$\xi'_-(s)=\frac{i\sqrt{-s}}{2}(f(s)-f^*(s^*)) \eqno(2.55)$$
instead of $\xi_-(s)$, putting on the functions
$\xi_+(s)$ and $\xi'_-(s)$ the requirement to decrease in the limit
$|s|\rightarrow\infty$
not slower then  $|s|^{-\alpha}$ and increased in the limit $|s|\rightarrow 0$
not faster then $|s|^{-1+\alpha}$. After repeating the calculations
performed earlier with the function $\xi'_-(s)$ instead of $\xi_-(s)$,
we get the following formula which replaces (2.43),
$$f'(s)=\frac{1}{4\pi}\int\limits_{0}^{\infty}ds'\frac{1}{s'-s}
\Bigl(\alpha(s',1)+\beta(s',1)-\alpha(s',-1)-\beta(s',-1)\Bigr)-$$
$$-\frac{i}{4\pi\sqrt{-s}}\int\limits_{0}^{\infty}ds'\frac{\sqrt{s'}}{(s'-s)}
\biggl(\Bigl(\alpha(s',1)-\beta(s',1)\Bigr)
+\Bigl(\alpha(s',-1)-\beta(s',-1)\Bigr)\biggr),\eqno(2.56)$$
where $f'(s)$ is an analogue of $f(s)$.
One sees, that in the cases, when properties of the
functions $\alpha$ and $\beta$
provide the existence of integrals in both formulas (2.43) and (2.56),
the following relation is true
$$f(s)-f'(s)=\frac{iC}{\sqrt{-s}},\eqno(2.57)$$
where $C$  is a real constant. Using the function $f'(s)$
instead of the function
$f(s)$  we get the equality
$$\theta(-k^2)(\alpha(k^2,0)-\beta(k^2,0))=-\theta(-k^2)
\frac{1}{2\pi\sqrt{-k^2}}\times$$
$$\times\int\limits_{0}^{\infty}ds'\frac{\sqrt{s'}}{s'-k^2}
\biggl(\Bigl(\alpha(s',1)-\beta(s',1)\Bigr)
+\Bigl(\alpha(s',-1)-\beta(s',-1)\Bigr)\biggr)\eqno(2.58)$$
instead of (2.45).
So, for the most natural restrictions imposed on the function $f(s)$,
which have under absence of massless particles the form
$$|f(s)|<|s|^{-1+\alpha_1}\quad\mbox{at}
\quad s\rightarrow 0 ,~\alpha_1>0,\eqno(2.59a)$$
$$|f(s)|<|s|^{-\alpha_2}\quad\mbox{at}
\quad s\rightarrow \infty ,~\alpha_2>0,\eqno(2.59b)$$
this function can be reconstructed from its imaginary
part above and below the cut
along the positive real axis with some arbitrariness.
This arbitrariness affects the relations like (2.45) and (2.58).
One can easily figure out the most general arbitrariness.
Let us assume that the functions $f_1(s)$ and $f_2(s)$
are analytical on the whole complex s - plane,
with the exception of the cut at $s\geq 0$, and that
they meet the condition
$$Im~f_1(s\pm i~\epsilon)=Im~f_2(s\pm
i~\epsilon)\quad\mbox{at}\quad s>0.\eqno(2.60)$$
We impose the condition (2.60) only at $s>0$, but not at $s=0$ in order to
consider specially a possible singularity at $s=0$. The function
$$\nu(s)\equiv f_1(s)-f_2(s)\eqno(2.61)$$
satisfies the equality
$$Im~\nu(s\pm~i\epsilon)=0,\quad\mbox{at}\quad s>0.\eqno(2.62)$$
Let us construct the functions
$$\chi_+(s)=\frac{1}{2}\Bigl(\nu(s)+\nu^*(s^*)\Bigr), ~~~~~\chi_-(s)=\frac{i\sqrt{-s}}{2}\Bigl(\nu(s)-\nu^*(s^*)\Bigr).\eqno(2.63)$$
Because of $(\sqrt{-s^*})^*=\sqrt{-s}$, one gets
$$\chi_{\pm}^*(s^*)=\pm\chi_{\pm}(s)\eqno(2.64)$$
and
$$Re~\chi_{\pm}(s+i~\epsilon)=Re~\chi_{\pm}
(s-i~\epsilon)\quad\mbox{at}\quad s>0. \eqno(2.65) $$
Further more due to (2.62)
$$Im~\chi_{\pm}(s+i~\epsilon)=0=Im~\chi_{\pm}
(s-i~\epsilon),\quad\mbox{at}\quad s>0.\eqno(2.66)$$
Consequently, at $s>0$
$$\chi_{\pm}(s+i~\epsilon)=\chi_{\pm}(s-i~\epsilon).\eqno(2.67)$$
Due to this fact the functions $\chi_{\pm}(s)$
are analytical in a whole complex $s$ -
plane
probably with the exception of singularity at $s=0$.
According to (2.61), (2.63)
$$\nu(s)\equiv f_1(s)-f_2(s)=\chi_+(s)-\frac{1}{\sqrt{s}}\chi_-(s). \eqno(2.68)$$
Due to analytical properties of functions $\chi_{\pm}$ under the conditions (2.59),
imposed on both quantities $f_1$ and $f_2$, there is only one possibility
\footnote{We abandon  simple proof of this fact for space saving. }
$$f_1(s)-f_2(s)=\frac{iC}{\sqrt{-s}},\quad
\mbox{where $C$ is a real constant}.\eqno(2.69)$$
We have already met this arbitrariness earlier (formula (2.57)).
So, under conditions (2.59) the  formula (2.69) describes all the arbitrariness
of restored function $f(s)$.

Further we shall be primarily interested in the case of the field
$\varphi$, annihilating the vacuum ($\varphi(x)|0\rangle=0$).
As already established,  this corresponds to the equality $\beta\equiv 0$.
Under such condition the relation (2.58) cannot take place for nonzero
functions $\alpha$, as far as all of this functions are nonnegative.
That's why we  previously used
the formula (2.43) but not (2.56), and we shall follow this assumption below.
Let us notice, in connection with this, that,
if we included the additional term (2.69)
in the right hand side of the formula (2.43),
we would not improve the u.v. behavior of the function $f$,
at positive $\alpha$ without disturbing a relation like (2.45).

Let us furthermore notice, that, weakening the condition (2.59a),
one could take into account massless particles, by including
into one of the functions $\alpha(k^2,+1)$, $ ~\beta(k^2,+1)$,
$~\alpha(k^2,-1)$, $~\beta(k^2,-1)$
an additional term $const\cdot\theta(k_0)\delta(k^2)$ or
correspondingly $const\cdot\theta(-k_0)\delta(k^2)$
and by using the formula (2.56). The relation (2.43)
in doing so is unapplicable, because of infinity
at $s\rightarrow 0$.  As it was just figured out, the formula (2.56) is incompatible
with the condition $\beta\equiv 0$. So we  see, that the Wightman
functions of the field $\varphi$ annihilating the vacuum
are not allowed to contain terms, which correspond
to massless particles. Further we assume, that there are no such particles,
and then we use the formula (2.43) and its consequences.

\section*{3.~~Constructing of free local fields with given propagators.}

In section 2 the general expression was deduced for the propagator of local
scalar field not forced to any restrictions
imposed on the spectrum of 4-momentum operator.
Generally speaking the field was not supposed to be free.
In order to use it further in perturbation theory we shall construct
now a free quantum field such that its propagator
coincides with the expression described in section 2.
Under the phrase "construct quantum field" we understand introducing the Hilbert space
of physical states and corresponding operators acting on it.

Obviously it suffices to construct free field, whose Wightman functions
in the momentum representation $\widetilde{u}(k)$
and $\widetilde{w}(k)$ coincide with preassigned ones,
since the propagator and  the commutator in this case,
can be restored from Wightman functions (see section 2). All assumptions and notations
in this section are the same as in section 2, but
we assume moreover that only two-point connected Green functions
of fields under consideration are nonzero.
We understand here the term "free fields" in this meaning only.
We choose a scheme of quantization which permits to describe fields with
different types of 4-momentum spectrum in a unique way.
Let us assume in the sake of simplicity
that  $\alpha$ and $\beta$ as functions of $k^2$
with fixed second argument $~\theta(k^2)~sgn(k_0)~$
are continuous everywhere probably with the exception of
no more than countable set of points and also can have
no more than countable set of singularities like
$c_1\delta(k^2-m^2),~ c_1 > 0$
\footnote{We exclude from consideration derivatives of $\delta$-
function because of its sign indeterminateness.}.
Unless otherwise stated the sign of $m^2$
is though here and below to be arbitrary.
The discrete
$\delta$-like singularities of $\alpha$ ($\beta$)
correspond to particles (antiparticles) with fixed square of mass.
Regions of continuity  of $\alpha$ and $\beta$
correspond to continues mass spectrum
i.e. "unparticle matter".

Further for arbitrary function  $y(x)$ we shall designate through
 $Supp~y$ the variety of its arguments $x$ for which $y(x)\neq 0$.

At first let us discuss the case when  $\delta$-like singularities are absent.
We assume below  that $\alpha$ and $\beta$
are defined and satisfy all conditions of section 2.
Furthermore we postulate that translation and Lorentz
invariant state $|0\rangle$ exists
and shall call it "vacuum".
Let us build Hilbert space as a Fock space upon the vacuum $|0\rangle$,
fixing the Lorentz frame of reference.
For this purpose let us introduce
operators $a(k)$ and $a^{\dag}(k)$ (correspondingly $b(k)$
and $b^{\dag}(k)$)for every $k\in Supp~\alpha$ ($~k\in Supp~\beta$),
which we call "annihilation and creation operators of particles with 4-momentum $k$"
(correspondingly antiparticles). Let us postulate that under Lorentz transformation
$a'(k')=a(k)$, $b'(k')=b(k)$.
Further let us introduce the conditions
$$a(k)|0\rangle =b(k)|0\rangle = 0\eqno(3.1)$$
and commutation relations
$$[a(k),a^{\dag}(k')]\equiv\frac{\delta^4(k-k')D_{\alpha}(k)}
{\alpha(k^2,\theta(k^2)~sgn(k_0))},\eqno(3.2)$$
where $D_{\alpha}(k)=1$, if $k\in Supp ~\alpha,$ and 0 otherwise.
In a similar manner for $b$ and $b^\dag$
$$[b(k),b^{\dag}(k')]\equiv\frac{\delta^4(k-k')D_{\beta}(k)}
{\beta(k^2,\theta(k^2)~sgn(k_0))}.\eqno(3.3)$$
All other commutators between creation and
annihilation operators are equal to zero.

Now we build
Fock space of ket vectors, acting on the vacuum by creation operators.
We assume that the operators $a^{\dag}(k)$ and $a(k)$
(correspondingly $b^{\dag}(k)$ and $b(k)$)
are Hermitian conjugated and
consequently
$\langle 0|a^{\dag}(k) = \langle 0|b^{\dag}(k)=0$.
This is a Hilbert space of states in our theory.

Further let us introduce local field $\varphi(x)$ in terms of creation and annihilation
operators by the formulas:
$$\varphi(x)\equiv\frac{1}{(2\pi)^2}\int d^4 k \{\alpha (k)e^{-ikx}a(k)+
\beta (k)e^{ikx}b^{\dag}(k)\},\eqno(3.4)$$
$$\varphi^{\dag}(x)\equiv\frac{1}{(2\pi)^2}\int d^4 k \{\alpha (k)e^{ikx}a^{\dag}(k)+
\beta (k)e^{-ikx}b(k)\}.\eqno(3.5)$$
Such field is obviously a scalar and satisfies the locality conditions from section 2,
if $\alpha$ and $\beta$
fulfill the relation (2.45).
Now let us build  4-momentum operator
$P_{\mu}$ for this field by means of creation and annihilation operators and
define it as follows
$$P_\mu\equiv\int d^4k~k_{\mu} \{\alpha(k)a^{\dag}(k)a(k)
+ \beta(k)b^{\dag}(k)b(k)\}.\eqno(3.6)$$
It is easy to see that standard relations take place:
$$\varphi(x)=e^{iPx}\varphi(0)e^{-iPx},
~~~~~P_{\mu}a^{\dag}(k)|0\rangle = k_\mu a^{\dag}(k)|0\rangle.\eqno(3.7)$$

Let us further describe a case of discrete mass spectrum.
For short
all relations are written for the particles creation and annihilation operators.
All formulas for antiparticles are similar.
Let us assume without loss of generality that $\alpha(k^2,\theta(k^2)~sgn(k_0))$
is nonzero only at  fixed value of its second argument and has only
one $\delta$- like singularity by its first argument.
Trying to write the relation (3.2) we have a difficulty
due to presence of $\delta(k^2-m^2)$ in the denominator in right hand side.
Therefore let us postulate the following relation that generalizes
(2) in the case of discrete spectrum
$$\alpha(k^2,\theta(k^2)~sgn(k_0))[a(k),a^{\dag}(k')]
\equiv \delta(k-k')D_{\alpha}(k).\eqno(3.8)$$
One can rewrite the equality (3.3) in a similar manner.
Relations (3.4), (3.5), (3.6), (3.7) remain the same.
Obviously, in the case of discrete mass spectrum, the argument $k$ of creation and
annihilation operators has only 3 independent components and we choose
its spacial components as independent ones.
So let us rewrite (3.8) in terms of operators that depend on $\vec{k}$ only.
Arbitrary function $\alpha(k^2,\theta(k^2)~sgn(k_0))$ can be presented in the form
$$\alpha(k) = \theta(k^2)\theta(k^0)\alpha(k^2,1)
+\theta(k^2)\theta(-k^0)\alpha(k^2,-1)
+\theta(-k^2)\alpha(k^2,0).\eqno(3.9)$$
Therefore, without loss of generality, it is enough to describe 3 cases only:
$$\alpha(k^2,1)\equiv\delta(k^2-m^2),
~~~~~~~\alpha(k^2,-1)\equiv 0,~~~~~~~
\alpha(k^2,0)\equiv 0 ,~~~~~m^2 > 0,\eqno(3.10)$$
$$\alpha(k^2,1)\equiv 0,~~~~~~~\alpha(k^2,-1)\equiv\delta(k^2-m^2),
~~~~~~~\alpha(k^2,0)\equiv 0 ,~~~~~~m^2 > 0,\eqno(3.11)$$
$$\alpha(k^2,1)\equiv 0,~~~~~~~\alpha(k^2,-1)\equiv 0,
~~~~~~~ \alpha(k^2,0)\equiv\delta(k^2-m^2),~~~~~m^2 < 0. \eqno(3.l2)$$

Let us notice that the case (3.10)
corresponds to the ordinary scalar field with mass $m$.
The case (3.11) corresponds to the scalar field with positive square of mass
but negative energy, and we call this field a "phantom".
The case (3.12) corresponds to a tachyon with fixed negative square of mass.
Of course, all spectrum should satisfy (2.45).
So in the case of tachyons, discrete parts of the functions
$\alpha$ and $\beta$ must coincide.

Let us notice that $\delta(k^2-m^2)$ with arbitrary sign of $m^2$
can be written in the form:
$\delta(k^2-m^2)=\frac{\theta(\vec{k^2}+m^2)}{2k^0}
\Bigl(\delta(k_0 - \sqrt{\vec{k}^2+m^2}) -
\delta(k_0 + \sqrt{\vec{k}^2+m^2})\Bigr)$.
At first we consider the case (3.10).
Let us integrate both parts of the relation (3.8) by $k^0$,
with the following result
$$[a(\vec{k}),a^{\dag}(\vec{k'})]=2k^0\delta^3(\vec{k}-\vec{k'}),
~ k^0=\sqrt{\vec{k}^2+m^2}, \eqno(3.13)$$
where $a(\vec{k})$ obviously is $a(k)$ at $~k_0=\sqrt{\vec{k}^2+m^2}$.
In the case (3.12) we get in the same way
$$[a(\vec{k}),a^{\dag}(\vec{k'})]=-2k^0\delta^3(\vec{k}-\vec{k'}),
~ k^0=-\sqrt{\vec{k}^2+m^2}. \eqno(3.14)$$
The case (3.12) is less trivial. In the fixed Lorentz frame of reference
let us introduce two sorts of annihilation operators
$a_{+}(\vec{k})\equiv a(k)$
at $~k_0=+\sqrt{\vec{k}^2+m^2} $ and $a_{-}(\vec{k})\equiv a(k)$
at $k_0=-\sqrt{\vec{k}^2+m^2}$ (correspondingly creation operators).
Let us notice that $a_+$ and $a_-$ transfer
into each other under Lorentz transformation
for some $k_{\mu}$.
This scheme is Lorentz invariant because
equalities  (3.4), (3.5), (3.6), (3.7), (3.8) are true.
Then the relation (3.9) leads to the following
commutation relations for $a_{\pm}(\vec{k})$ and
$a_{\pm}^{\dag}(\vec{k})$:
$$[a_+(\vec{k}),a_+^{\dag}(\vec{k'})]=2k^0\delta^3(\vec{k}-\vec{k'})
\theta(\vec{k}^2+m^2)\quad\mbox{at}\quad k^0=\sqrt{\vec{k}^2+m^2}, \eqno(3.15a)$$
$$[a_-(\vec{k}),a_-^{\dag}(\vec{k'})]=-2k^0\delta^3(\vec{k}-\vec{k'})
\theta(\vec{k}^2+m^2)\quad\mbox{at}\quad k^0=-\sqrt{\vec{k}^2+m^2}. \eqno(3.15b)$$
For example let us rewrite the formula (3.4) in the case (3.12) as follows
\footnote{This case corresponds to tachyons with fixed square of mass, and
these tachyons must be $CPT$-invariant i.e. $\alpha\equiv\beta$ (see (2.45)).}:
$$\varphi(x)=\int \frac{d^3\vec{k}~\theta(\vec{k}^2+m^2)}{2\sqrt{\vec{k}^2+m^2}}
\biggl(\Big(e^{-i\sqrt{\vec{k}^2+m^2}x_0+i\vec{k}
\vec{x}}a_{+}(\vec{k})+e^{+i\sqrt{\vec{k}^2+m^2}x_0
+i\vec{k}\vec{x}}a_{-}(\vec{k})\Big)+$$
$$+\Big(e^{+i\sqrt{\vec{k}^2+m^2}x_0-i\vec{k}\vec{x}}b^{\dag}_{+}(\vec{k})+
e^{-i\sqrt{\vec{k}^2+m^2}x_0-i\vec{k}\vec{x}}
b_{-}^{\dag}(\vec{k})\Big)\biggr).\eqno(3.16)$$
Each term including  $a_{+}$ or $a_{-}$
(correspondingly $b^{\dag_{+}}$ or $b^\dag_{-}$)
is not Lorentz invariant but their sum is, because
it can be presented as the first term
(correspondingly the second) of the expression (3.4).
Let us remark, that in the example (3.16) the commutator
$[ \varphi(x), \varphi^{\dag}(y) ]$ is identically equal to zero,
but the Wightman functions and the propagator are nonzero and the latter
has the following unconventional form:
$G(k) = i\delta(k^2-m^2)$ (see (2.50)).

Finally let us consider a question about
classical action and how it should be quantized in order
to get the local free quantum field $\varphi(x)$ described above.
We shall built this field $\varphi$ as a sum with certain coefficients
of nonlocal fields with fixed
square of mass, and shall quantize  the last ones  Lorentz invariantly in a way
which leads to the formulas (3.4) and (3.5).

In the following let the index $j$ run through values  $0,~+1,~-1$.
For every $m^2\in Supp~\alpha(m^2,j)$
let us introduce classical scalar nonhermitian field
with fixed square of a mass $\varphi_{m^2,~\alpha,~j}(x)$.
In the same manner we define $\varphi_{m^2,~\beta,~j}(x)$
for every $m^2\in Supp~\beta(m^2,j)$.
At this stage there are no principal differences
in a properties  of this fields but they will be quantized in a different ways.
Let us construct an action of a usual type for each of the fields introduced above
$$S_{m^2,~n,~j}\equiv\int d^4x ~\varphi^{\dag}_{m^2,~n,~j}(x)\left(\partial_{\mu}
\partial^{\mu}+m^2\right)\varphi_{m^2,~n,~j}(x),\eqno (3.17)$$
where index $n$ ranges over $\alpha$, $\beta$.

As far as each of the fields $\varphi_{m^2,~n,j}(x)$ satisfies the
corresponding  Klein-Fock-Gordon equation,  the following equalities are true
$$\varphi_{m^2,~n,~j}(x)=\frac{1}{(2\pi)^2}\int d^4 k ~\delta (k^2-m^2)
\widetilde{\varphi}_{m^2,~n,~j}(k)e^{-ikx},\eqno (3.18)$$
$$\varphi^{\dag}_{m^2,~n,~j}(x)=\frac{1}{(2\pi)^2}\int d^4 k ~\delta (k^2-m^2)
\widetilde{\varphi^*}_{m^2,~n,~j}(k)e^{+ikx}.\eqno (3.19)$$
In the expressions (3.18) and (3.19) the functions
$\widetilde{\varphi}(k)$ and $\widetilde{\varphi^*}(k)$
\footnote{We do not write sub-indexes for short.}
are arbitrary scalar functions defined on
$Supp~\alpha$ and $Supp~\beta$ (depending on the value of the index "n").
Now let us postulate the following \textbf{\emph{nonstandard}} recipe
of quantization, i.e. rules
by which one transforms classical fields, written in the form
(3.18) and (3.19) into operators acting
on Hilbert space introduced above. These operators satisfy
commutation relations (3.2),(3.3)
(or (3.13),(3.14),(3.15) for discrete mass spectrum. The rules look as follows:
$$\widetilde{\varphi}_{m^2,~\alpha,~+1}(k)
\rightarrow a(k)\theta(+k^2)\theta(+k_0),~~~~~
\widetilde{\varphi^*}_{m^2,~\alpha,~+1}(k)
\rightarrow a^{\dag}(k)\theta(+k^2)\theta(+k_0),$$
$$\widetilde{\varphi}_{m^2,~\alpha,~-1}(k)
\rightarrow a(k)\theta(+k^2)\theta(-k_0),~~~~~
\widetilde{\varphi^*}_{m^2,~\beta,~-1}(k)
\rightarrow a^{\dag}(k)\theta(+k^2)\theta(-k_0),$$
$$\widetilde{\varphi}_{m^2,~\alpha,~0}(k)
\rightarrow a(k)\theta(-k^2),~~~~~
\widetilde{\varphi^*}_{m^2,~\beta,~0}(k)
\rightarrow a^{\dag}(k)\theta(-k^2).$$
There are similar equalities for the $b$ and $b^{\dag}$.
Let us notice that quantum fields $\varphi_{m^2,~n,~j}(x)$ and ones
conjugated with them,
do not satisfy locality conditions.
Now, using the operators just defined, we construct
local field  $\varphi(x)$ according to the formula:
$$\varphi(x)\equiv\int\limits_{-\infty}^{+\infty} d m^2
\sum\limits_{j=-1,~0,+1}\{\alpha(m^2,j)
\varphi_{m^2,~\alpha,~ j}(x)+\beta(m^2,j)
\varphi^{\dag}_{m^2,~\beta,~j}(x)\}.\eqno(3.20)$$
One can built the field $\varphi^{\dag}(x)$ in a similar manner.
It is easy to see that our last constructions coincide with
formally defined ones by equalities (3.4) and (3.5).

Let us emphasize that in contrast to usual theory where
commutators between creation and annihilation operators
follow from canonical commutation relations between
generalized coordinates and momenta,
in this scheme we deduced them from the requirement of getting the given
Wightman functions. The Fourier transforms of these functions
satisfy all limitations from section
2 and in particular the formula (2.45).  Finally one
gets free local scalar fields defined by
(3.4) and (3.5) and the free Hamiltonian equal to $P_0$ from the equality (3.6).
In the sectors of particles with positive square
of mass and positive or negative energy,
this two approaches
give the same results.

Having such free fields one can introduce an interaction of them
with other fields and investigate it in the interaction picture,
drawing Feynman diagrams.

\section*{4.~~Local relativistic generalization of Lindblad equation.}
One of the most interesting applications of local field annihilating the vacuum
consists in using it for relativistic generalization of Lindblad equation for
density matrix $\rho$.
It is known that in  nonrelativistic Quantum Mechanics this
equation has the form [7]\footnote{See also [8].}:
$$\frac{d\rho}{dt}=-i[H,\rho]+\sum\limits_n \lambda_n\left(2A_n\rho A^{\dag}_n
-A^{\dag}_n A_n\rho-\rho A^{\dag}_n A\right),\eqno(4.1)$$
where  $H$  is usual Hamiltonian, $A_n$  are  arbitrary operators, $\lambda_n$
are positive constants.

Let us consider the case of one nonhermitian scalar field in a flat
space-time and require Lorentz invariance.  To establish the latter in the simplest way
one may use the Tomonaga-Schwinger formalism [9][10] in the interaction picture.
We assume that the operators
$\varphi(x)$ form  in this picture a free local field and that the
density matrix $\rho(\sigma)$ describing the state depends on
space-like hypersurface $\sigma$. This density matrix is defined
for each such surface. Along with the field
$\varphi(x)$ there may exist other fields and
their state is described by $\rho(\sigma)$ too. Then one can write the following
relativistic generalization of Lindblad equation
$$\frac{\delta\rho(\sigma)}{\delta\sigma(x)}=-i\left[H_1(x),\rho(\sigma)\right]
+\lambda\left(2\varphi(x)\rho\varphi^{\dag}(x) -
\varphi^{\dag}(x)\varphi(x)\rho-\rho\varphi^{\dag}(x)\varphi(x)\right),\eqno(4.2)$$
where $\frac{\delta\rho(\sigma)}{\delta\sigma(x)}$ is  variational derivative
of $\rho(\sigma)$ induced by infinitely small change $\delta\sigma(x)$ of
the surface $\sigma$ in the vicinity of the point $x$, $H_1(x)$ is the part of
hamiltonian density describing the interaction of the field
$\varphi(x)$ with itself and with other fields as well as of other fields
with themselves in the interaction picture.
The $\lambda$ is  positive constant,
$\delta \sigma (x)$ is the 4-space volume enclosed between
space-like surface $\sigma$ and the varied surface
$\sigma+\delta\sigma(x)$.
Resolvability condition of the equation (4.2) is
$$\frac{\delta^2\rho(\sigma)}{\delta\sigma(x_1)\delta\sigma(x_2)}
=\frac{\delta^2\rho(\sigma)}{\delta\sigma(x_2)\delta\sigma(x_1)},\eqno(4.3)$$
and it may be fulfilled under local
commutativity of the operators $\varphi(x)$,
$\varphi^{\dag}(x)$
and $H_1(x)$.In other words, if $\left(x_1-x_2\right)^2 < 0$,
the following conditions must hold:
$$\left[\varphi(x_1),\varphi(x_2)\right]=0,~~~
\left[\varphi(x_1),\varphi^{\dag}(x_2)\right]=0,~~~
\left[\varphi(x_1),H_1(x)\right]=0. \eqno(4.4)$$
All other fields presented in our system are thought to be local.
Let us show that for local scalar field
$\varphi(x)$ nonannihilating the vacuum the equation (4.2)
leads to irremovable u.v. divergences.
For the sake of simplicity consider the case when there are no other fields
with the exception of the
$\varphi$ and when $H_1=0$. Let us assume, in particular, that
on the surface $\sigma$ the state $\rho(\sigma)$
is the  vacuum of the interaction picture: $\rho(\sigma)=|0\rangle\langle0|$.
Under the variation $\delta\sigma(x)$ of the surface
$\sigma$ in vicinity of the point $x$ it appears that
$\rho(\sigma+\delta \sigma)=|0\rangle\langle0|
+\left(\frac{\delta\rho(\sigma)}{\delta\sigma(x)}\right)\delta\sigma(x)$,
where $\frac{\delta\rho(\sigma)}{\delta\sigma(x)}$
is defined by the equation (4.2). Let us ask the question: what is the probability
of the vacuum state $|0\rangle\langle0|$ to remain unchanged?
To answer this question one should calculate
$Sp\left\{|0\rangle\langle 0|\rho\left(\sigma+\delta\sigma(x)\right)\right\}=
      \langle 0|\rho\left(\sigma+\delta\sigma(x)\right)|0\rangle    $.

As much as $\varphi$ is in the interaction picture with the vacuum $|0\rangle$,
we assume that
$\langle 0|\varphi(x)| 0\rangle =\langle 0|\varphi^{\dag}(x)
| 0\rangle = 0 $ at every $x$.
So we have:
$$\langle 0|\rho\left(\sigma+\delta\sigma(x)\right)|0\rangle
=1-2\lambda\langle 0|\varphi^{\dag}(x)\varphi(x)
|0\rangle\delta\sigma(x)+O\left(\left(\delta\sigma(x)\right)^2\right).\eqno(4.5)$$
The $\langle 0|\varphi^{\dag}(x)\varphi(x)|0\rangle$
is the Wightman function of the fields  $\varphi(x)$ and $\varphi^{\dag}(x)$
at coinciding arguments. Even in the general case considered
above in section 2 this is equal to $+\infty$,
if it is not identically zero.
The last possibility corresponds to the field annihilating the vacuum.
Indeed, according with (2.6), (2.8), (2.16)
$$\langle 0|\varphi^{\dag}(x)\varphi(x)
|0\rangle=\frac{1}{(2\pi)^4}\int d^4 k \beta(k^2,\theta(k^2)~sgn(k_0)).\eqno(4.6)$$
Because $\beta\geq 0$ one sees by going to the hyperbolic
coordinates that this expression is proportional
to the infinite volume of the hyperboloid $\theta(\pm k_0)(k^2-1)=0$,
or correspondingly $k^2+1 = 0$, unless $\beta \equiv 0$.
Let us also notice that the mentioned divergence is the divergence of probability
but not of probability amplitude. So due to positivity condition
we are not allowed to subtract any counterterms
to renormalize it.

The considered problem was discussed by Pearl [2],
who suggested to refuse local commutativity
of the field $\varphi$ and consequently to renounce the proof of Lorentz invariance
using an equation like (4.2). Under this conditions
the Lorentz invariance  must be proved directly
by passage from one Lorentz frame to another.
But even in this case it appears to be necessary to introduce
tachyonic field.

The described above problem of u.v.
divergence connected with an equation like (4.2),
does not arise, if the field $\varphi(x)$
annihilates the vacuum state, i.e. if  $\beta\equiv 0$.
If simultaneously $H_1=0$, then due to the equation (4.2)
the vacuum state $\rho=|0\rangle\langle0|$
will not change with the course of time at all.
Consequently this is the only possibility, accordant
with the local field theory.
This possibility should be investigated.

As found out above the field $\varphi(x)$
annihilating the vacuum can exist only
in a theory with continuous tachyonic spectrum.
If under absence of Lindblad- terms
(i.e at $\lambda=0$) this field is free, then it belongs to the class
of generalized free fields and its pure tachyonic part describes
"unparticle matter".
The existence of such field might be
assumed without contradiction with experimental
data  supposing only, that it interacts with other matter very weakly.
Just the possibility to apply such a field in relativistic Lindblad equation
impelled the authors to undertake the present investigation.
To make a theory with the field $\varphi$ practically useful
one has to introduce
very weak interaction of this field with usual matter.
As it will be figured out below we meet very hard difficulties on this way and
so far we did not remove them.
Now let us describe this problem.

\section*{5.~~Difficulties caused by interaction between the
field $\varphi$ annihilating the vacuum and
ordinary fields. }

Consider now a practically important case when
the system described by Lindblad equation (4.2) contains
ordinary fields interacting with the field $\varphi$.
We assume for simplicity that there is only
one ordinary hermitian scalar field $\xi(x)$ with the mass $m$, $m^2>0$.
The Hamiltonian density $H_{int}(x)$ in the Lindblad equation (4.2)
describes the interaction between the fields $\varphi(x)$,
$\varphi^{\dag}(x)$ and $\xi(x)$ only. All the fields are thought
in the interaction picture. Before the interaction is taken
into account the following formula holds
$$\varphi(x)|0\rangle=0,\eqno(5.1)$$
Let us require that the interaction does not destroy the property
(1) of the vacuum state. This means that the vacuum in the $Schr\ddot{o}dinger$ picture
coincides with the one in the interaction picture. It leads to the condition
$$H_{int}(x)|0\rangle=0.\eqno(5.2)$$
Without this requirement it is impossible to introduce
a pure state which is stable in time and
fulfills the Lindblad equation, i.e. it is impossible
to define the vacuum state correctly
for this equation.

The simplest density of interaction Hamiltonian, fulfilling the condition (5.2)
looks like this
$H_{int}(x)=g\xi(x)\varphi^{\dag}(x)\varphi(x).$
All operators written here and below are in
the interaction picture. One can develop invariant
perturbation theory and draw Feynman diagrams for this theory.
Free propagator $D_\varphi(k)$ of the field $\varphi(x)$
in the momentum representation is defined by the formula (2.53).
We shall represent it on diagrams by an arrowed line
(because $\varphi$ is nonhermitian).
The propagator of the field $\xi(x)$  equals to
$D_\xi(k)=\frac{1}{m^2-k^2-i\epsilon}.$
We shall represent it on diagrams using wavy line.
Let us show that this theory is nonrenormalizeble.
Consider the diagram with 5 tales shown in fig 1.

  \includegraphics[width=6cm]{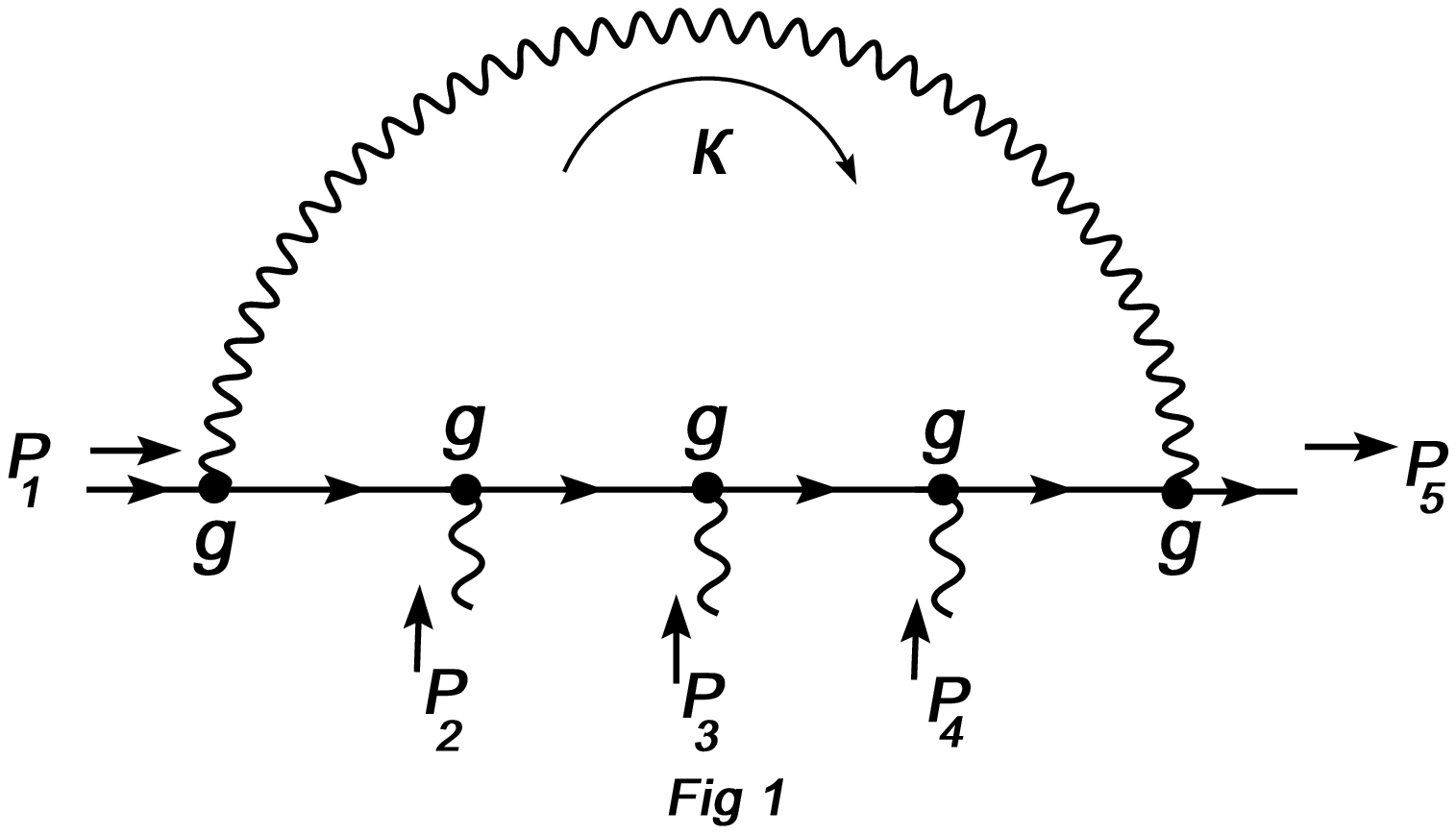}

We shall show in a moment that although the difference between momentum
degrees in the numerator and in the denominator is negative, this
diagram has u.v. divergence.
The last fact is closely connected with special analytical structure
of the $\varphi$ field propagator. let us beforehand notice that the divergence
of such diagram leads to nonrenormalizability of considered theory.
Really, the presence of this divergence forces us to
introduce a vertex of 5-th degree in fields during renormalization:
$g_5\xi^3(x)\varphi^{\dag}(x)\varphi(x).$
It is easy to see that canonical dimensions of the fields  $\varphi$ and
$\xi$ are equal to $+1$ (see the expressions for the propagators), and
hence the canonical dimension of the $g_5$ is $-1$.
Therefore the theory
is nonrenormalizable. Indeed, in this case
there is a new divergence  connected with 7- tale diagram shown in fig 2.
Consequently one needs to introduce
 an additional term
$g_7\xi^5(x)\varphi^{\dag}(x)\varphi(x)$
in the $H_{int}$. But this leads to a new
divergence (see fig 3) and so on.
Finally we conclude that our theory is nonrenormalizable.

\includegraphics[width=5.5cm]{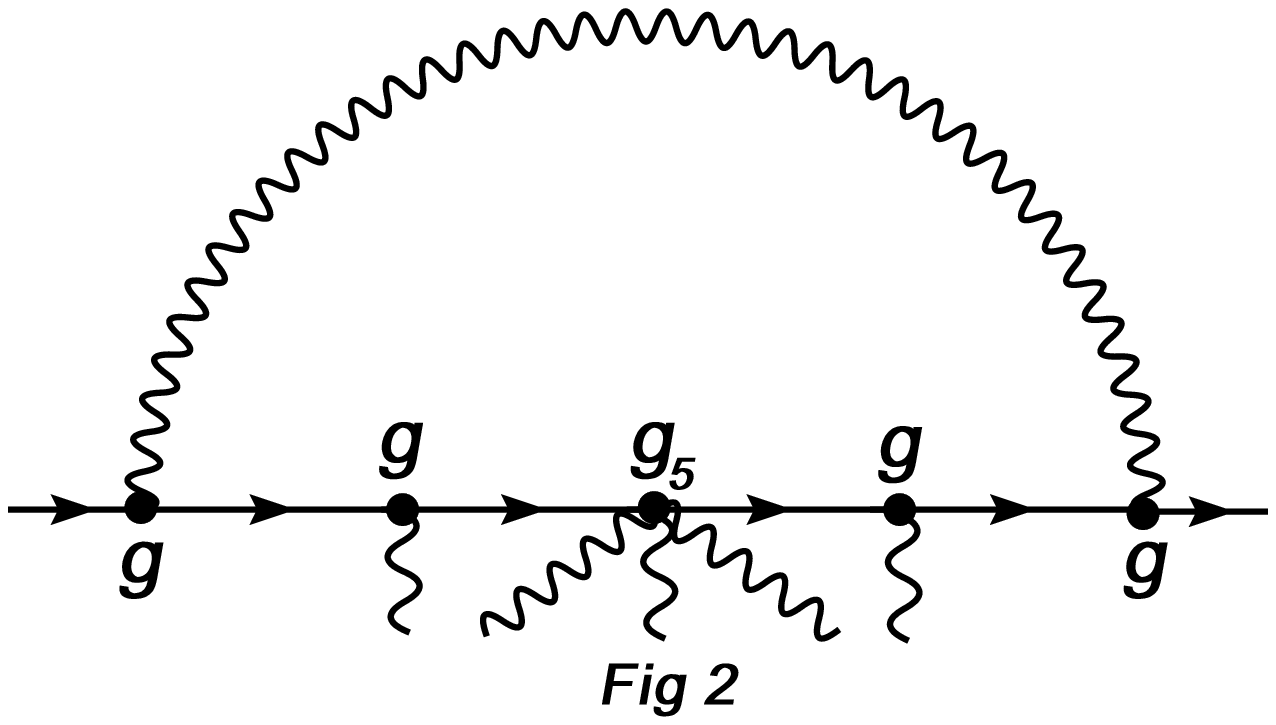}
\includegraphics[width=5.5cm]{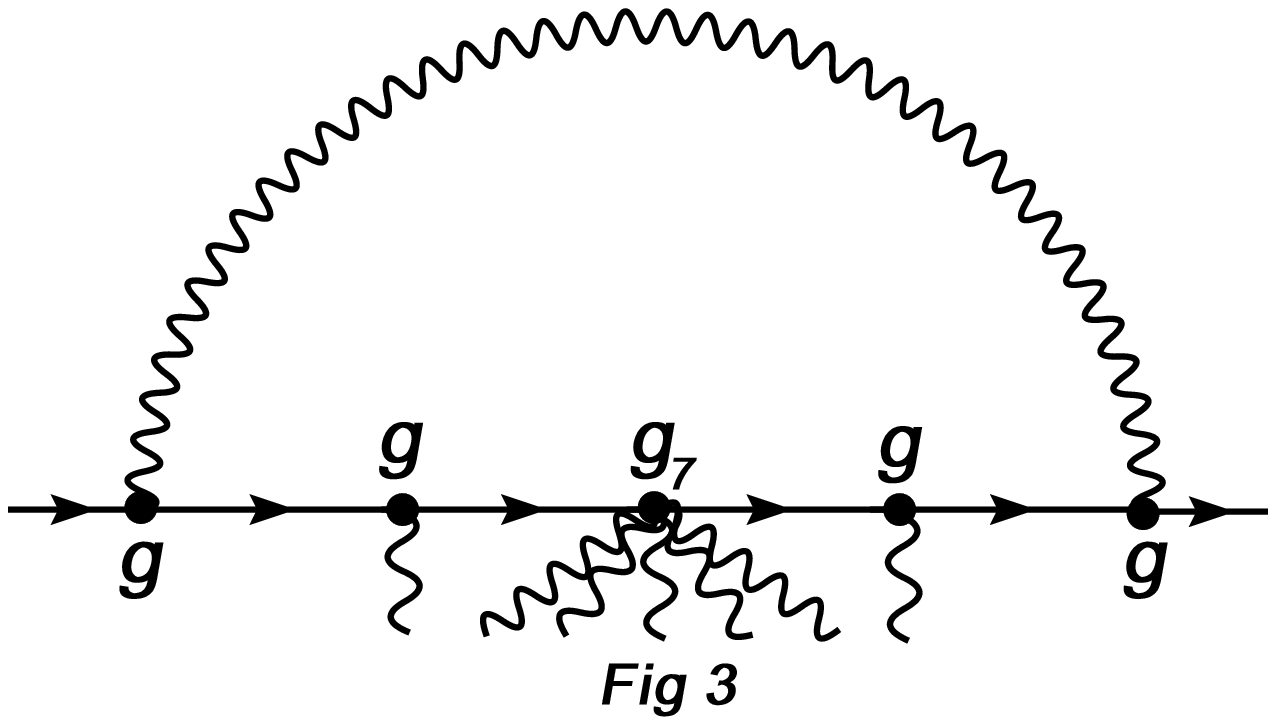}

So let us now return to the diagram in fig 1.
At first, we shall consider a special case of the (2.53) for which
$$\alpha(k^2,1)=\delta(k^2-m_1^2)\quad\alpha(k^2,-1)\equiv 0
\quad\mbox{where}\quad m^2_{1}>0. \eqno(5.3)$$
So one should calculate the following integral
$$\int d^4k D_\xi(k)D_\varphi(p_1-k)D_\varphi(p_1+p_2-k)
D_\varphi(p_1+p_2+p_3-k)D_\varphi(p_1+...+p_4-k).\eqno(5.4)$$
From the formula (2.53)
one sees, that the propagator $D_\varphi$ as a function of $k^0$ at fixed $\vec{k}$,
has two branch points  at  $k_0 = \pm |\vec{k}|$
and two poles in the points $k_0=\pm\sqrt{(\vec{k})^2+m_1^2}$.
Due to presence of "$i\epsilon~sgn(k^0)$" in the denominator
and under the sign of the square root,
the contour of integration by $k^0$ passes above this singularities
in complex plane of the variable $k^0$.
In turn, the singularities by $k^0$
of the $\xi$ field propagator are passed over by the contour of integration
in the following manner: $k^0=-\sqrt{\vec{k}^2+m^2}$
 - from below,
and $k^0=+\sqrt{\vec{k^2}+m^2}$ - from above.
This implies that in contrast with the ordinary theory
we are not allowed to make  "Euclidian rotation"
in the complex plane of the variable $k^0$.
On the other hand, we can perform the integration
over $k^0$ using the theorem of residues, by closing
path of integration around the upper complex semiplane.
So, the quantity (4) turns into
$$\left.\pi i\int\frac{d^3\vec{k}}{\sqrt{\vec{k}^2+m^2}}
\Bigr(D_\varphi(p_1-k)\cdot...
\cdot D_\varphi(p_1+...+p_4-k)\Bigr)\right|_{k^0=-\sqrt{\vec{k}^2+m^2}}.\eqno(5.5)$$
Let us consider the behavior
of the expression under integral sign in the
(5.5) at $|\vec{k}|\rightarrow\infty$.
There are 4 propagators $D_\varphi$ in the fig 1;
let us number them in the following way:
the momentum $p_1+...+p_j-k~~(j=1,2,3,4)$
flows through the propagator of number $j$.
From the equality (2.53) and
our additional assumption (5.3) one can see that
anyway at $p^0_1+...+p^0_j > |\vec{p}_1+...+\vec{p}_j| $, the
$j$-th propagator $D_{\varphi}$ decreases not faster than
$\frac{const(j,p_1,...,p_4)}{\sqrt{|\vec{k}|}}$,
when $|\vec{k}|\rightarrow+\infty$. So, their product
decreases not faster than $\frac{1}{|\vec{k}|^2}$ and
the expression under integral sign
as a whole - not faster than $\frac{1}{|\vec{k}|^3}$, hence the integral diverges.
We would get the same result, if we consider a case when
$\alpha(k^2,-1)=\delta(k^2-m_2^2),~~m_2^2>0$.
Presenting nonnegative continuous functions $\alpha(k^2,\pm 1)$ in the form
$\alpha(k^2,\pm 1)=\int\limits_0^{+\infty}dm^2\delta(k^2-m^2)\alpha(m^2,\pm 1)$,
we make sure that the most general expression (2.53)
does not make the situation better.

Let us notice that similar result takes place if we use arbitrary
scalar field considered in the section 2 and nonannihilating vacuum as
the  field $\xi$.
Really, the propagator of just considered ordinary field $\xi$
can be written in the form (compare with (2.48a)):
$$\frac{1}{m^2-k^2-i\epsilon} =
\frac{1}{m^2-k^2-i\epsilon~sgn(k^0)}+2\pi i \theta(-k_0)\delta(k^2-m^2)\eqno(5.6)$$
In a discourse made above only the second term of the last expression
contributed to our diagram (as integration contour passed above both
singularities of the first term).
If we consider the similar diagram taking the expression
$2\pi i \theta(-k_0)\delta(k^2-m^2)$
as a wavy line,
then the result will be obviously the same.
Meanwhile, it is easy to see that the sign of $k^0$ and the sign of $m^2$
are not important for our discourse
i.e. if we calculate  diagram shown in fig 1 and use the expression
$\theta(\pm sgn (k_0))\delta(k^2-m^2)$ at arbitrary sign of $m^2$ as wavy line,
it  still diverges.
Finally let us look at the formula (2.48 a).
Only the second term "$i\beta(k^2,-\theta(k^2)~sgn(k^0))$" (compare with (5.6))
gives contribution to the diagram similar to the one, presented in fig 1.
Let us present $\beta$ in the form:
$$\beta(k)=\theta(k^0)\beta_1(k^2)+\theta(-k^0)\beta_2(k^2)=$$
$$=\int\limits_{-\infty}^{+\infty}dm^2\beta_1(m^2)\theta(k^0)\delta(k^2-m^2)+
\int\limits_{-\infty}^{+\infty}dm^2\beta_2(m^2)\theta(-k^0)\delta(k^2-m^2),$$
where $\beta_{1,2}\geq 0$. One can see that u.v.
divergence, similar to that one described above,
takes places every time with the exception of the case $\beta\equiv 0$
that corresponds to the field
$\xi$ annihilating the vacuum.

So we  conclude that under condition (5.2) the field $\varphi(x)$
is not able to interact with the ordinary fields in  renormalizable way.
If we took vector or spinor
field as a field of matter and made discourse
like one written above, we would have
the same result.
Moreover, if we try to introduce the interaction between matter fields and
the field  $\varphi$
not directly but through other local scalar
fields of a general form described in the section 2,
we have nonrenormalizability again.

Everywhere above we required
that the interaction of the field
$\varphi$, subordinated to the condition $\varphi|0\rangle = 0$,
with other fields does not disturb the stability of the vacuum $|0\rangle$.
As a matter of principle it is possible that this stability
is only approximate but the rebuilding of the vacuum flows
extremely slowly. Then one may assume that
at the birth of the Universe, the vacuum
fulfilled the condition $\varphi|0\rangle = 0$
and it did not change significantly till nowadays.
We did not investigate this possibility in full scale, but
after consideration of
several variants we found out that the rebuilding of the vacuum in these
examples is infinitely rapid.

\subsection*{Conclusion.}
As a result of our investigation the generalization of
Kallen-Lehmann representation for the propagator of local
scalar field with arbitrary spectrum of 4-momentum operator
is established. Under this conditions the $CPT-$ violation becomes possible.
The 4-momentum spectrum of $CPT-$violating field is not arbitrary
but obeys the relation like (2.45) which follows from the locality condition.
Furthermore the tachyons with discrete mass spectrum are not allowed
to violate $CPT-$invariance.
It is found out that at this assumption and under $CPT$- violation
nonzero local field annihilating the vacuum state can exist. Such field
is composite: it may have terms which correspond to positive square
of mass and positive or negative sign of energy, but it must contain
\emph{continues} tachyonic spectrum of mass.

Using such field we succeeded to write a correct
(in particular without u.v. divergences) local relativistic
generalization of Lindblad equation for statistical
operator in the case of no interaction with other fields.

It turned out that for the simplest interaction
of described field with ordinary ones, the theory becomes
nonrenormalizable.
We investigated other simple examples of interaction of
local tachyonic and phantomic fields
\footnote{We remind that we call the field a "phantom"
if its square of mass is positive, but the energy is negative.}
(in particular, nonannihilating
the vacuum) with ordinary ones.
These examples, not described in this work, also
did not lead to renormalizable theories. Such results are
closely connected with the general form of propagator (2.50)
and with impossibility to make "Euclidian rotation".

In the light of the consideration made above,
the question arises, whether
the local tachyonic and phantomic fields can
interact in a renormalizable way with the ordinary fields at all.
In the case of positive answer only, the $CPT$- noninvariant
local fields can interact with ordinary fields in a
renormalizable theory.
At the present time we do not know the answer.

\end{document}